\documentclass[12pt,preprint]{aastex}

\begin{document}

\title{Contamination of cluster radio sources in the measurement of 
the thermal Sunyaev-Zel'dovich angular power spectrum}

\author{Wei Zhou and Xiang-Ping Wu}

\affil{National Astronomical Observatories, Chinese Academy
                 of Sciences, Beijing 100012, China}

\begin{abstract}
We present a quantitative estimate of the confusion of 
cluster radio halos and galaxies in the measurement of  
the angular power spectrum of the thermal Sunyaev-Zel'dovich (SZ) effect.
To achieve the goal, we use a purely analytic approach
to both radio sources and dark matter of clusters by incorporating
empirical models and observational facts together with 
some theoretical considerations. It is shown that 
the correction of cluster radio halos and galaxies
to the measurement of the thermal SZ angular power spectrum is 
no more than $20\%$ at $l>2000$ for observing frequencies 
$\nu>30$ GHz. This eliminates the concern that the SZ 
measurement may be seriously contaminated by the existence of 
cluster radio sources. 
\end{abstract}


\section{Introduction}

As a result of gravitationally driven shocks and compression,  
majority of the baryons in the most massive, collapsed objects in the
universe such as groups and clusters of galaxies exist in the form
of hot plasma with temperature of $10^{7}-10^{8}$ K. 
These hot baryons manifest themselves by strong diffuse X-ray sources 
in the sky within the framework of bremsstrahlung emission,
which are directly detectable with current X-ray instruments. 
On the other hand,  the hot electrons 
in the virialized massive dark halos also scatter the passing
cosmic microwave background (CMB) photons, giving rise to a subtle
change in the CMB spectrum,  known as the Sunyaev-Zel'dovich (SZ) effect 
(Sunyaev \& Zel'dovich 1972). 
Unlike the X-ray emission of the hot gas in massive halos, which
shows a strong dependence on the detailed structures and enrichness history 
of the hot gas as well as the distances of the host halos from us, 
the SZ effect is completely determined by the total thermal energy 
of the gas intrinsic to the systems. Therefore, it is well
suited for studies of distant, massive clusters, which may provide 
a stringent constraint on cosmological parameters especially for the
mean matter density $\Omega_{\rm M}$ and the normalization of 
matter power spectrum $\sigma_8$ (e.g. Barbosa et al. 1996;
Molnar, Birkinshaw \& Mushotzky 2002; Komatsu \& Seljak 2002). 
A more practical and powerful utilization of the
SZ effect nowadays is perhaps to statistically measure the angular 
power spectrum of the SZ signals at various scales.  This 
allows us to gain the weak SZ signals at high statistical significance
levels without resolving individual groups and clusters. 
Indeed, recent detection of the excess power relative to
primordial CMB anisotropy at arcminute scales has been attributed to
the statistical signals of the thermal SZ effect
(Dawson et al. 2001; Bond et al. 2003).  A number of ongoing and
future experiments will soon be able to provide a more robust measurement 
of the SZ angular power spectrum at high multipoles or small angular scales.

There are many other effects that should be corrected for before 
the SZ angular power spectrum is used for cosmological purposes.
These include nongravitational effect on the distribution and
evolution of the hot gas inside halos (da Silva et al. 2000, 2001;
Seljak, Burwell \& Pen 2001; Xue \& Wu 2001; 
Holder \& Carlstrom 2001; Zhang, Pen \& Wang, 2002; Komatsu \& Seljak 2002; 
Zhang \& Wu 2003; etc.), possible asymmetric shape  
of dark halos (Cooray 2000; Lee \& Suto 2003; 
Piffaretti, Jetzer \& Schindler 2003), 
contaminations of radio point sources 
(Cooray et al. 1998; Lin, Chiueh \& Wu 2002; Holder 2002; 
Rubin{\~o}-Mart\'in \& Sunyaev 2002), etc.   
Despite large uncertainties, many attempts have been made 
over the past few years towards the evaluation of these effects 
on the SZ cluster counts and angular power spectrum measurements.  
In this work, we will explore the contamination of radio sources in 
clusters in the measurement of the thermal SZ angular power spectrum,
namely, cluster radio halos and galaxies.

In theory of hierarchical clustering, massive dark halos like 
clusters of galaxies form by gravitational aggregation of individual 
low-mass objects. Sub-halo mergers will generate a considerably large
number of energetic electrons that can be accelerated by 
cluster magnetic fields over cluster regions, giving rise to the radio 
halos due to synchrotron emission (Buote 2001 and reference therein).
Regardless of the debate on the acceleration mechanism for
relativistic electrons in clusters, about $1/3$ of nearby, rich clusters 
exhibit radio halos, and this fraction may increase with 
redshift if mergers were more frequent in the past
(En{\ss}lin \& R\"ottgering 2002). 
Furthermore, there is a strong observational evidence that
cluster radio halos show a
morphology similar to the diffuse X-ray emission from the hot
intracluster gas, and the corresponding radio power is correlated 
strongly with the X-ray luminosity and temperature of clusters 
(Liang et al. 2000;  Govoni et al. 2001a,b). This enables
one to estimate the distribution and evolution of cluster radio
halos based on the well-determined X-ray properties of clusters. 
For example, one can easily derive the radio luminosity function of 
clusters at different wavebands and redshifts 
(En{\ss}lin \& R\"ottgering 2002).
Together with the radio surface brightness profiles of clusters
motivated by empirical formula (e.g. $\beta$ model) for 
the X-ray surface brightness distribution of clusters, 
one can estimate the radio background from the diffuse radio halos of
clusters. 

Existence of diffuse radio halos constitutes a major source of 
contaminations in the measurement of the thermal SZ effect toward
clusters. Since radio halos always contribute a positive signal
superimposed on the CMB sky, the decrements of the CMB signals 
due to the thermal SZ effect of 
the hot intracluster gas at low frequencies would be compensated, 
while the increments of the SZ signals at high frequencies will be
correspondingly enhanced. The crossover point occurs at $\nu=217$ GHz.
Moreover, the amplitude of such a modification to the thermal
SZ signals depends critically on observing frequencies.  
This arises because the radio flux of 
cluster halos as a result of synchrotron emission has a steep spectrum 
of roughly $S_{\nu}\propto \nu^{-1}$.  Consequently, 
at very high frequencies the
cluster radio halos may become hardly visible, and no 
modification to the SZ signals from the cluster radio halos
is needed. However, with the decrease of observing frequencies  
the SZ decrements and the diffuse radio emission of cluster halos 
may appear to be comparable, and eventually the latter would 
become the dominant component in the microwave sky. This will lead to 
a nonnegligible correction to the SZ measurement (En{\ss}lin 2002)
especially in the central regions of clusters, because
the radio emission is more centrally concentrated than the
thermal SZ effect if the radio emission is assumed to follow the 
X-ray emission.  Therefore, it deserves a close investigation of 
the influence of the cluster radio halos on the measurement of 
the SZ angular power spectrum.

Another source of contaminations in the SZ measurements 
is radio point sources, which include radio galaxies
inside clusters and tracing large-scale structures of the universe.
In a similar way to cluster radio halos, cluster radio galaxies
also compensate the decrements of the thermal SZ intensity in the 
Rayleigh-Jeans region. On the other hand, the Poisson distribution
and clustering of radio galaxies on large scales would 
enhance the radio fluctuations, resulting in
a positive contribution to both primary CMB and SZ power spectra. 
The role of these field radio galaxies has been well addressed 
by several authors (e.g. Franceschini et al. 1989; 
Toffolatti et al. 1998), and the resulting contamination in 
the measurement of CMB anisotropies is found to be generally small 
or can be somehow removed.    
Nevertheless, a significant fraction of the SZ signals could be 
diluted by the unsubtracted radio sources inside clusters
(Cooray et al. 1998; Lin, Chiueh \& Wu 2002).   
Using a toy model of radio point source subtraction, 
in which a certain central region is removed from 
all clusters, Holder (2002) showed that the true SZ power 
at $l>1000$ can be underestimated by as much as $30\%$ at 
$\nu=30$ GHz.  A more realistic model for the radial distribution 
of radio galaxies in clusters is thus needed to refine 
this estimation.

In this paper, using a purely analytical approach, we perform 
a quantitative study of the contamination of cluster radio halos
and galaxies in the measurement of the angular power spectrum 
of the thermal SZ effect. 
It is hoped that such a study may clarify the issues of 
to what extent one has to take the correction of cluster radio sources 
into account in the cosmological applications of the thermal SZ angular 
power spectrum.
Throughout the paper we will work with a flat cosmological model of 
$\Omega_{\rm M}=0.35$ and $\Omega_{\Lambda}=0.65$.

\section{Thermal SZ angular power spectrum}

The change in CMB temperature $T_{\rm CMB}$ in direction 
${\mbox{\boldmath $\theta$}}$ due to the thermal SZ effect of 
the hot electrons in massive clusters is simply
\begin{eqnarray}
\frac{\Delta T({\mbox{\boldmath $\theta$}})}{T_{\rm CMB}} &=& 
           g(x)\;y\left({\mbox{\boldmath $\theta$}}\right);\\
y(\theta)&=&\int n_{\rm e}\sigma_{\rm T} \left
(\frac{kT_{\rm e}}{m_{\rm e}c^2}\right) d\chi;\\
g(x) &=& x\frac{e^x+1}{e^x-1}-4,
\end{eqnarray}
where $x=h_{\rm p}\nu/kT_{\rm CMB}$, $n_{\rm e}$ and $T_{\rm e}$ are the
number density and temperature of electrons in clusters, respectively,
and the integral is performed along the line of sight $\chi$.
The angular power spectrum of the temperature 
fluctuation $\Delta T({\mbox{\boldmath $\theta$}})/T_{\rm CMB}$
can be separated into the Poisson term $C_{\ell}^P$ and 
the clustering term $C_{\ell}^C$:
\begin{equation}
C_{\ell} = C_{\ell}^P+C_{\ell}^C.
\end{equation}
Since the clustering term $C_{\ell}^C$ is several orders of magnitude
smaller than the Poisson term $C_{\ell}^P$ at small angular 
scales $\ell \geq 1000$ where the SZ power becomes dominant 
in the CMB sky, we will neglect the contribution of source 
clustering to the SZ angular power spectrum. 
In the flat sky approximation, the Poisson term can be calculated by
(e.g. Cole \& Kaiser 1988)
\begin{eqnarray}
C_{\ell}^P = g^2(x)\int dz\frac{d^2V}{dzd\Omega}
        \int dM\frac{d^2N(M,z)}{dMdV}|y_{\ell}(M,z)|^2,
\end{eqnarray}
in which $d^2V/dzd\Omega$ denotes the comoving volume per unit redshift and
per steradian, and $y_{\ell}(M,z)$ is the Fourier transform of 
$y(\theta)$:
\begin{equation}
y_{\ell} = 2\pi \int y(\theta)J_0(\ell\theta)\theta d\theta, 
\end{equation}
where $J_0$ is the Bessel function of zero order. 
We adopt the standard Press \& Schechter (1974) formalism
for the mass function of clusters   
\begin{equation}
\frac{d^2N}{dMdV}=-\sqrt{\frac{2}{\pi}} \frac{\bar{\rho}}{M}
    \frac{\delta_{\rm c}(z)}{\sigma^2(M)}
    \frac{d\sigma(M)}{dM}
    \exp{\left(-\frac{\delta_c^2(z)}{2\sigma^2(M)} \right)},
  \label{eq:ps}
\end{equation}
where $\bar{\rho}$ is the mean mass density of the universe, 
$\delta_c(z)$ is the linear 
over-density of spherical collapse that virialized at redshift $z$, and
$\sigma(M)$ is the present rms mass fluctuation within the 
top-hat filter $M=4\pi\bar{\rho}R^3/3$:
\begin{equation}
\sigma^2(M)=\frac{1}{2\pi^2}\int_0^{\infty}k^2P(k)W^2(kR)dk,
\end{equation}
in which $W(x)=3(\sin x-x\cos x)/x^3$ is the Fourier
representation of the window function. We parameterize the power
spectrum of fluctuation $P(k)\propto k^{n_{\rm s}}T^2(k)$, and take the fit
given by Bardeen et al. (1986) for the transfer function of
adiabatic CDM model $T(k)$ with the shape parameter $\Gamma=0.21$. 
The primordial power spectrum is assumed
to be the Harrison-Zel'dovich case $n_{\rm s}=1$. The amplitude in the power
spectrum is fixed using the rms fluctuation
on an 8 $h^{-1}$ Mpc scale, $\sigma_8$, which will be taken to be 
$\sigma_8=0.9$ in this study.

\section{Contribution of cluster radio halos}
 
\subsection{Properties of cluster radio halos}

Over the past few years, diffuse radio halos have been detected 
in a few tens of nearby, rich clusters. They often extend to a 
distance of $\sim1$ Mpc from cluster centers, 
and have regular shape, low surface brightness 
and steep radio spectrum $S_{\nu}\propto \nu^{-1}$. 
Moreover, observations suggest that there is a tight correlation 
between the radio power $P_{1.4{\rm GHz}}$ and the X-ray 
luminosity ($L_{\rm X}$) of host clusters
(Liang et al. 2000; Govoni et al. 2001b), which can be well described 
by the following fitting formula (En{\ss}lin \& R\"{o}ttgering 2002)
\begin{equation}
P_{1.4{\rm GHz}} = a_{\nu}10^{24}\;h_{50}^{-2}\;{\rm Watt\,Hz}^{-1}\;
            \left(\frac{L_{\rm X}}{10^{45}\;h_{50}^{-2}
                  {\rm erg\;s}^{-1}}\right)^{b_\nu},
\end{equation}
in which $a_{\nu} = 2.78$, $b_\nu = 1.94$, and $h_{50}$ is the present
Hubble constant in units of $50$ km s$^{-1}$ Mpc$^{-1}$. 
In order to set up a link between the X-ray luminosity and 
total mass of the cluster, we first
employ the best-fit $L_{\rm X}$-$T$ relation of Xue \& Wu (2000), 
based on the hitherto largest sample of nearby X-ray clusters, 
\begin{equation}
\left(\frac{L_{\rm X}}{10^{43}\,h_{50}^{-2}\;{\rm erg\,s}^{-1}}\right)=
        10^{-0.032}\left(\frac{kT}{\rm keV}\right)^{2.79}.
\end{equation}
It has been shown that such an $L_{\rm X}$-$T$ relation  
exhibits no significant evolution at least out to $z\sim0.8$ 
(Rosati, Borgani \& Norman 2002 and references therein).
We then convert the X-ray temperature into the total cluster mass 
using virial theorem (Bryan \& Norman 1998):
\begin{equation}
\frac{kT}{kT^*} =\left(\frac{M}{10^{15}M_\odot}\right)^{2/3},
\end{equation}
in which 
\begin{equation}
kT^*=1.39\,{\rm keV}\;f_{\rm T}[h^2\Delta_{\rm c}(z)E^2(z)]^{1/3},
\end{equation}
$h$ is the Hubble constant in units of 100 km s$^{-1}$ Mpc$^{-1}$,
$f_{\rm T}$ is the normalization factor, which will be taken to be
$f_{\rm T}=0.8$ in this work, 
and $\Delta_{\rm c}(z)$ is the overdensity of the 
virialized dark halos with respect to the critical density of the
universe at redshift $z$. Equivalently, one may also employ the
empirical $L_{\rm X}$-$M$ relation (Reiprich \&  B\"{o}hringer 1999)
incorporating with a simple model of cosmic evolution for dark halos.  
Consequently, for a given halo of mass $M$
one is able to estimate its radio power at 1.4 GHz through
\begin{equation}
P_{1.4{\rm GHz}} = 3.17\times 10^{20}\,h_{50}^{-2}\,{\rm Watt\, Hz}^{-1}
     \left(\frac{kT^*}{\rm keV}\right)^{5.41}
     \left(\frac{M}{10^{15}M_\odot}\right)^{3.61}.
\end{equation}
Radio observations indicate that cluster radio halos usually have 
a steep power index, which is also well motivated by synchrotron 
emission process. We adopt a spectral 
power index of  $\alpha = 0.8$ (e.g. Giovannini et al. 1993) to convert
the radio power at 1.4 GHz to the radio emission at other wavelengths.

Furthermore, several observations have also reveled a similarity of the radio 
and X-ray morphologies in clusters (Deiss et al. 1997; Liang et al. 2000;
Govoni et al. 2001a; Feretti et al. 2001). This enables us to model the
radio surface brightness of clusters using the empirical fitting formula
for X-ray surface brightness profile, i.e., the $\beta$ model,
\begin{equation}
\Delta I_{\rm RH}(r) = \Delta I_{\rm RH0}
       \left(1+\frac{r^2}{r_{\rm c}^2}\right)^{-3\beta+0.5},
\end{equation}
in which the normalization factor can be determined by the total radio
power $P_{\lambda}$ at a given wavelength $\lambda$. Meanwhile,  
the electron number density in clusters follows
\begin{equation}
n_{\rm e}(r) = n_{\rm e0}\left(1+\frac{r^2}{r_{\rm c}^2}\right)^{-3\beta/2}.
\end{equation}
Here we introduce the universal baryon fraction 
$f_{\rm b}=\Omega_{\rm b}/\Omega_{\rm M}$ to
fix the central electron density $n_{\rm e0}$: $f_{\rm b}=M_{\rm gas}/M$, 
where $M_{\rm gas}$ is the total gas mass within $r_{\rm vir}$.
The gaseous halo (also the radio halo) is truncated at the virial radius 
$r_{\rm vir}$ defined by 
\begin{equation}
M=\frac{4\pi}{3}r_{\rm vir}^3\Delta_{\rm c}(z)\rho_{\rm c}(z),
\end{equation}
where $\rho_{\rm c}(z)$ is the critical mass density of the universe at $z$.
We adopt a constant value of $2/3$ for the $\beta$ parameter, and 
assume that the core radius $r_{\rm c}$ of gas distribution is proportional
to the size of the host halo: $r_{\rm c}\propto r_{\rm vir}$. We specify the
proportionality constant such that $r_{\rm c}=0.15$ $h^{-1}$ Mpc 
for  a massive cluster of $M=10^{15}h^{-1}M_{\odot}$ 
at $z=0$ (cf. Komatsu \& Kitayama 1999).

Finally, radio halos are preferentially found in clusters that demonstrate 
violent mergers and substructures, which provide an accelerating 
mechanism for the relativistic particles responsible for radio emission
(e.g. Buote 2001).  Yet, the fraction of clusters that possess giant 
radio halos is still uncertain. Following En{\ss}lin \& 
R\"{o}ttgering (2002), we
adopt a constant radio halo fraction of $f_{\rm RH}=1/3$. Nevertheless,
we will also demonstrate the result for a fraction of unity for comparison.  
Recall that cluster mergers may occur more frequently in the past
than at the present.

\subsection{Modification to the SZ angular power spectrum}

In the Rayleigh-Jeans region, the presence of radio halos can 
dilute the thermal SZ signals towards clusters. The significance
of the effect depends critically on working frequencies.
Recall that radio emission of clusters has a steep power 
spectrum of $\alpha\approx0.8$.
At very low frequencies, the decrements of the CMB 
temperature due to the SZ effect could be 
completely compensated (En{\ss}lin 2002) by cluster radio
halos. Here we demonstrate the radio halo contamination
using an observing frequency of $\nu=30$ GHz.
The change in the CMB temperature is related to the intensity 
fluctuations by
\begin{equation}
\frac{\Delta T}{T_{\rm CMB}} = \frac{(e^x-1)^2}{x^4e^x}\frac{\Delta {I}}{I_0},
\end{equation}
where $I_0=2(kT_{\rm CMB})^3/(h_{\rm p}c)^2$. $\Delta I$ is composed 
of two components, the thermal SZ signals $\Delta I_{\rm SZ}$ and
the cluster radio surface brightness  $\Delta I_{\rm RH}$,
\begin{equation}
\Delta I = \Delta I_{\rm SZ} + \Delta I_{\rm RH},
\end{equation}
or equivalently, we can separated the temperature fluctuation into 
\begin{equation}
\frac{\Delta T}{T_{\rm CMB}}= 
         \frac{\Delta T_{\rm SZ}}{T_{\rm CMB}} + 
         \frac{\Delta T_{\rm RH}}{T_{\rm CMB}}.
\end{equation}
If the fraction of cluster radio halos is $f_{\rm RH}$, then
the observed SZ power spectrum at small angular scales is 
\begin{equation}
C_{\ell} = \int dz\frac{d^2V}{dzd\Omega}\int dM\frac{d^2N(M,z)}{dMdV}
           \left[(1-f_{\rm RH})|g(x)y_{\ell}(M,z)|^2+
                f_{\rm RH} |y_{\ell}^{\prime}(M,z)|^2\right],
\end{equation}
where $y_{\ell}^{\prime}(M,z)$ is the Fourier transform of 
$(\Delta T_{\rm SZ}+\Delta T_{\rm RH})/T_{\rm CMB}$.

In order to highlight the effect of cluster radio halos on the
SZ angular power spectrum,  we demonstrate in Figure 1 the 
relative correction to the thermal SZ angular power spectrum 
$|\Delta C_{\ell}|/C_{\ell}$, along with the 
contributions of clusters in different redshift ranges.  
At $1000\la\ell\la 8000$, the radio halo contamination in the 
SZ angular power spectrum arises mainly from clusters at
$0.1<z<0.5$.  High-redshift clusters come into effect only at 
very small angular scales $\ell>10^4$, while large angular scales are 
dominated by low-redshift clusters. 
In Figure 2 we show the contributions of cluster radio halos with
different cluster masses to the thermal SZ angular power spectrum.
At large angular scales below $\ell\approx500$, the radio contamination
is governed by nearby, massive clusters, which can be easily removed 
from the SZ map because of their rarity. At smaller angular scales beyond
$\ell=1000$, the contamination arises mainly from moderately 
rich clusters at higher redshifts $z>0.2$. 
Contributions from poor clusters and groups of masses less than 
$2\times10^{14}$ $M_{\odot}$ are negligible because of the absence 
of radio halos in these systems.  To provide a quantitative 
illustration of the modifications due to cluster radio halos,
we list in Table 1 the relative corrections $|\Delta C_l|/C_l$
for a set of $\ell$ evaluated by Equation (20),
assuming the radio halo fractions of
$f_{\rm RH}=1/3$ and $f_{\rm RH}=1$, respectively. 
We have also shown in Table 1 how our estimates are affected by 
the uncertainties associated with the measurements of 
radio power, X-ray luminosities and total masses of clusters.
To do this, we simply vary the radio power $P_{\lambda}$, 
X-ray luminosity $L_X$ and normalization factor $f_T$ in the
virial theorem by $20\%$, $20\%$ and $30\%$, respectively, 
which are roughly representative of the uncertainties in current 
determinations of these quantities.  
It turns out that the modifications to the 
thermal SZ angular power spectrum from cluster radio halos 
are only minor, and the maximum correction is no more than 
$13\%$ for $\ell>2000$ even in the extreme case of where $f_{\rm RH}=1$
and $f_T=1$. Major uncertainty in our calculation arises from
the mass estimate of clusters characterized by the normalization 
factor $f_T$, which could differ by a factor of $\sim2$ among different 
simulations and observations (e.g. Bryan \& Norman 1998; 
Huterer \& White 2002).

\begin{table*}
 \vskip 0.2truein
\caption{Relative corrections to the SZ angular power spectrum from
cluster radio halos}
 \vskip 0.2truein
\small
\begin{tabular}{cccccc}
\tableline 
\tableline
$f_{\rm RH}$ & 
      $\ell=2000$  & $\ell=3000$  & $\ell=4000$ & $\ell=5000$\\ 
\tableline
 $1/3$   & $1.15^{+0.09+0.11+3.10}_{-0.14-0.29-1.03}\%^*$   & 
           $0.84^{+0.08+0.11+2.18}_{-0.12-0.23-0.81}\%$   & 
           $0.69^{+0.07+0.11+1.48}_{-0.11-0.21-0.70}\%$   & 
           $0.58^{+0.07+0.12+0.93}_{-0.10-0.20-0.63}\%$  \\ 
\tableline
  1      & $3.77^{+0.26+0.34+8.64}_{-0.43-0.88-3.09}\%$   & 
           $2.92^{+0.23+0.32+5.78}_{-0.36-0.71-2.45}\%$   & 
           $2.46^{+0.22+0.34+3.64}_{-0.32-0.64-2.11}\%$   & 
           $2.26^{+0.22+0.36+2.01}_{-0.30-0.59-1.89}\%$ \\
\tableline
\end{tabular}

 \parbox {6.in}{$^{*}$ Listed in order are uncertainties arising from
                 $20\%$, $20\%$ and $30\%$ variations in radio power,
                 X-ray luminosity and mass estimate of clusters,
                 respectively.}
\end{table*}

\section{Contribution of cluster radio galaxies}

\subsection{Distribution of cluster radio galaxies}

A critical issue in the estimate of confusion of cluster radio
galaxies in the measurement of the SZ angular power spectrum is to 
properly handle the distribution of cluster radio galaxies.  
This includes the luminosity function and radial profile of 
radio galaxies in clusters.  We adopt the univariate radio luminosity
function given by Ledlow \& Owen (1996), based on a statistical complete
sample of 188 radio galaxies in Abell clusters within $z=0.09$ and at
$\nu=1.4$ GHz. The univariate radio luminosity function 
$f_{1.4{\rm GHz}}$ can be well fitted by a broken power law 
\begin{equation}
\log f_{1.4{\rm GHz}}=\log \frac{dN_{\rm radio}/d\log P_{1.4{\rm GHz}}}
             {N_{\rm gal}(0.3r_{\rm Abell})}
             =a+b\log P_{1.4{\rm GHz}},
\end{equation}
where $N_{\rm gal}(0.3r_{\rm Abell})$ is the total number of galaxies 
brighter than $R=-20.5$ within 0.3 Abell radius of the cluster center, 
and $(a,b)$ are proportionality constants
with ($-0.150,1.766$) and ($-1.433,33.667$) for 
$P_{1.4{\rm GHz}}<10^{24.8}$ W\,Hz$^{-1}$ and 
$P_{1.4{\rm GHz}}>10^{24.8}$ W\,Hz$^{-1}$,
respectively. 
In order to convert the above radio luminosity function at 
$\nu=1.4$ GHz into the ones ($f_{\nu}$) at other frequencies, 
we use the spectral index distribution fitted by 
Lin, Chiueh, \& Wu (2002) for a sample of 64 radio sources 
in 56 clusters observed by Cooray et al. (1998), 
in which the mean spectral index is 
$\alpha=0.71$.  This allows us to take the dispersion of 
spectral index into account. The luminosity-weighted cluster 
radio luminosity function takes the form of 
\begin{equation}
\log{(P_{\rm \nu}f_{\rm \nu})}=\left\{
\begin{array}{ll}
C_{\rm 1}(\nu)+\gamma_{\rm 1}(\nu)\log{P_{\rm \nu}}; & \log{P}<\log{P_{\rm b}}
\\
C_{\rm 2}(\nu)+\gamma_{\rm 2}(\nu)\log{P_{\rm \nu}}; &
\log{P}>\log{P_{\rm b}}
\end{array}\right.,
\end{equation}
where
\begin{eqnarray}
C_{\rm 1}(\nu)&=&2.68-0.95e^{-0.015\nu}  \nonumber \\
\gamma_{\rm 1}(\nu)&=&0.85e^{-0.0027\nu}+1.49\times10^{-3}\nu \nonumber\\
C_{\rm 2}(\nu)&=&30.42+3.29e^{-0.025\nu}\nonumber \\
\gamma_{\rm 2}(\nu)&=&-0.43e^{-0.0022\nu}-6.14\times10^{-4}\nu,
\end{eqnarray}
and the break power $P_{\rm b}$ has a weak dependence on frequency: 
log$P_{\rm b}(\nu) = 23.86 -0.00396\nu$.

We assume that the number density profile of galaxies in clusters 
follows the King model
\begin{equation}
n_{\rm gal}(r) = n_{\rm gal}(0)\left(1+\frac{r^2}{r_c^2}\right)^{-3/2}.
\end{equation}
The surface number density is thus 
$\Sigma_{\rm gal}(r)=\int n_{\rm gal}(r) d\chi$. 
A good approximation of the core radius is $r_c=0.1r_{\rm vir}$. 
In order to fix the central density $n_{\rm gal}(0)$, we introduce
the so-called halo occupation distribution, 
$N_{\rm gal}(M)=\int n_{\rm gal}(r)4\pi r^2dr$. 
We adopt the best fit analytic formulae of Sheth \& Diaferio (2001) 
for the spiral [$N_{\rm gal,S}(M)$] and elliptical [$N_{\rm gal,E}(M)$]
galaxies based on the GIF simulations (Kauffmann et al. 1999):
\begin{eqnarray}
N_{\rm gal,S}(M)&=&(M/M_{\rm S})^{\alpha_S}+
                   0.5e^{-4[\log(M/10^{11.75}M_{\sun})]^2} \nonumber\\
N_{\rm gal,E}(M)&=&(M/M_{\rm E})^{\alpha_E}
                      e^{-(2\times10^{11}M_{\sun}/M)^2} \nonumber\\
N_{\rm gal}(M)&=&N_{\rm gal,S}(M)+N_{\rm gal,E}(M), 
\end{eqnarray}
where $M_{\rm S}=7\times10^{13}h^{-1}M_{\sun}$, $\alpha_{\rm S}=0.9$,
$M_{\rm E}=3\times10^{12}h^{-1}M_{\sun}$, and $\alpha_{\rm E}=0.75$.

Another parameter we need to fix before we proceed to our numerical
calculation is $N_{\rm gal}(0.3r_{\rm Abell})$, the galaxies 
brighter than $R=-20.5$ within 0.3 Abell radius $r_{\rm Abell}$.
Instead of evaluating $N_{\rm gal}(0.3r_{\rm Abell})$ directly,
we turn to the fraction of galaxies with absolute magnitude 
brighter than $R_{\rm limit}=-20.5$, $f_{\rm gal}$. This can be achieved by 
employing the luminosity functions of galaxies in clusters, 
$\Phi_S$ and $\Phi_E$. We adopt the best fit Schechter 
forms in the $r^*$ band for early and late type galaxies derived from
the Sloan Digital Sky Survey (Goto et al. 2002). 
The best fit parameters in Goto et al. (2002) are based on 
galaxies within $0.75$ $h_{0.7}^{-1}$ Mpc, which is only slightly 
larger than $0.3r_{\rm Abell}$.  We convert the 
$r^*$ magnitude to the standard Johnson-Morgan-Cousins system
using $r^*=R+0.16(V-R)+0.13$ (Fukugita et al. 1996) and then 
assign the colors to early and late type galaxies according to 
the tabulated values given by Fukugita, Shimasaku \& Ichikawa (1995). 
Moreover, we choose a lower magnitude cutoff of $R_{\rm cut}=-18.5$
to guarantee the validity of the above adopted halo occupation 
distribution equation (25) which accounts for galaxies brighter than 
$V=-17.7+5\log h$. Taking all these factors into account, we have  
\begin{eqnarray}
f_{\rm gal}=\frac{\int_{r^*_{\rm S,\,limit}}^{\infty} \Phi_S(r^*)dr^* 
             +\int_{r^*_{\rm E,\,limit}}^{\infty} \Phi_E(r^*)dr^*}
             {\int_{r^*_{\rm S,\,cut}}^{\infty} \Phi_S(r^*)dr^* 
             +\int_{r^*_{\rm E,\,cut}}^{\infty} \Phi_E(r^*)dr^*}.
\end{eqnarray}

The total radio power of all galaxies in a cluster can be obtained 
by extrapolating the radio luminosity function of Ledlow \& Owen (1996) 
determined with  $0.3r_{\rm Abell}$ to the whole cluster region out 
to virial radius, which yields 
\begin{eqnarray}
P_{\rm tot,\nu} &= &
      f_{\rm gal}\left[\frac{N_{\rm gal}(M)}
           {N_{\rm gal}(0.3r_{\rm Abell})} \right] 
       \int P_{\nu}   (dN_{\rm radio}/d\log P_{\nu}) d\log P_{\nu} \nonumber\\
   &=& f_{\rm gal} N_{\rm gal}(M) \int P_{\nu} f_{\nu} d\log P_{\nu}.
\end{eqnarray}
The total surface brightness of all radio galaxies measured at 
frequency $\nu$ in direction  ${\mbox{\boldmath $\theta$}}$ toward 
a cluster of mass $M$ and redshift $z$ is thus 
\begin{eqnarray}
\Delta I_{\rm RG}({\mbox{\boldmath $\theta$}}) &=& 
	\left[\frac{\Sigma_{\rm gal}({\mbox{\boldmath $\theta$}})D_A^2(z)}
             {N_{\rm gal}(M)}\right]
         \left[\frac{(1+z)P_{\rm tot,\nu}}{4\pi D_L^2(z)} \right] \nonumber\\
   &=& 
    \frac{\Sigma_{\rm gal}({\mbox{\boldmath $\theta$}})f_{\rm gal}
    \int P_{\rm \nu}f_{\rm\nu}d{\rm log}P_{\rm \nu}}{4\pi(1+z)^3},
\end{eqnarray}
where $D_A$ and $D_L$ are the angular diameter distance and luminosity 
distance to the cluster, respectively. 
This adds to the SZ signals such that the observed radio intensity is
$\Delta I = \Delta I_{\rm SZ} + \Delta I_{\rm RG}$. 
The corresponding angular power spectrum reads 
\begin{equation}
C_{\ell} = \int dz\frac{d^2V}{dzd\Omega}\int dM\frac{d^2N(M,z)}{dMdV}
                |y_{\ell}^{\prime \prime}(M,z)|^2,
\end{equation}
where $y_{\ell}^{\prime \prime}(M,z)$ is the Fourier transform of 
$(\Delta T_{\rm SZ} + \Delta T_{\rm RG})/T_{\rm CMB}$, and the 
temperature fluctuation $\Delta T_{\rm RG}$ is related to 
$\Delta I_{\rm RG}$ following equation (17).

\subsection{Modification to the SZ angular power spectrum}

In a similar way to Figure 1, we plot in Figure 3 
the relative corrections to the SZ angular power spectrum 
$|\Delta C_{\ell}|/C_{\ell}$ because of the presence of radio galaxies 
in clusters, together with the contributions of cluster radio 
galaxies in different redshift ranges. At $l>1000$, the distant 
radio galaxies at $z>0.2$, which could be regarded
as the unresolved radio sources, dominate the correction term.
We list in Table 2 the relative corrections for a set of 
small scales, together with the uncertainties corresponding to
$20\%$ variation of the radio luminosity function. 
It turns out that the confusion of unresolved radio galaxies 
in clusters in the measurement of the thermal SZ effect is less 
than $7.5\%$ on all scales beyond $l>2000$. This is  
somewhat smaller than the estimate ($30\%$) of Holder (2002),
who evaluated the effect of radio point sources by removing a 
certain central region ($\sim1^{\prime}$) from all clusters. The latter
could simultaneously lead to an underestimate of the thermal SZ effect
in the centers of all clusters especially for high-redshift clusters.

\begin{table*}
 \vskip 0.2truein
\caption{Relative corrections to the SZ angular power spectrum from
cluster radio galaxies}
 \vskip 0.2truein
\begin{tabular}{cccccc}
\tableline 
\tableline
redshift & 
      $\ell=2000$  & $\ell=3000$  & $\ell=4000$ & $\ell=5000$\\ 
\tableline
 $0.2<z<5$  & $4.39^{+0.86}_{-0.88}\%$ & 
              $4.63^{+0.90}_{-0.92}\%$ & 
              $4.69^{+0.90}_{-0.92}\%$ & 
              $4.69^{+0.89}_{-0.92}\%$\\
\tableline
 $0<z<5$  &   $6.47^{+1.07}_{-1.16}\%$ & 
              $5.80^{+0.97}_{-1.05}\%$ & 
              $5.41^{+0.91}_{-0.98}\%$ & 
              $5.16^{+0.87}_{-0.94}\%$\\
\tableline
\end{tabular}
\end{table*}

\section{Conclusions}

Contamination of radio sources has been a major concern in 
the measurement of thermal SZ effect, which include  
radio halos and galaxies of clusters, and clustering of
radio galaxies tracing large-scale structures of the 
universe. The latter comes into effect only on large angular 
scales and appears to be relatively small as compared with the thermal SZ
effect (Toffolatti et al. 1998). We have thus concentrated 
in this paper on the role of cluster radio halos and galaxies. 
The effect of cluster radio halos on the measurement of the SZ angular
power spectrum has previously remained unknown, while the confusion of 
the cluster radio galaxies has been recently shown to be 
very significant (Holder 2002). Using an analytic model for
both radio sources and dark matter of clusters, 
we have performed a quantitative analysis of the extent to which 
cluster radio halos and galaxies may contaminate the measurement of 
thermal SZ angular power spectrum. We have found that while
the cluster radio sources can indeed cause some  
confusion in the measurement of SZ angular power spectrum, 
the relative correction $|\Delta C_l|/C_l$
is well within $20\%$ at $l>2000$ for an observing frequency of
$\nu=30$ GHz. The modification becomes even smaller for high-frequency 
observations. 
We thus conclude that there is probably no need to take 
the confusion of cluster radio sources into account 
in the thermal SZ measurement, if our observing frequencies are 
chosen to be $\nu>30$ GHz, unless the thermal SZ power spectrum is
used for the purpose of precision cosmology.

\acknowledgments
We gratefully acknowledge Torsten En{\ss}lin for useful discussion and help,
and an anonymous referee for useful suggestions. 
This work was supported by the National Science Foundation of China, 
under Grant No. 10233040, and the Ministry of Science and Technology of 
China, under Grant No. NKBRSF G19990754.

\clearpage


\begin{figure}
\epsscale{0.8}   
\plotone{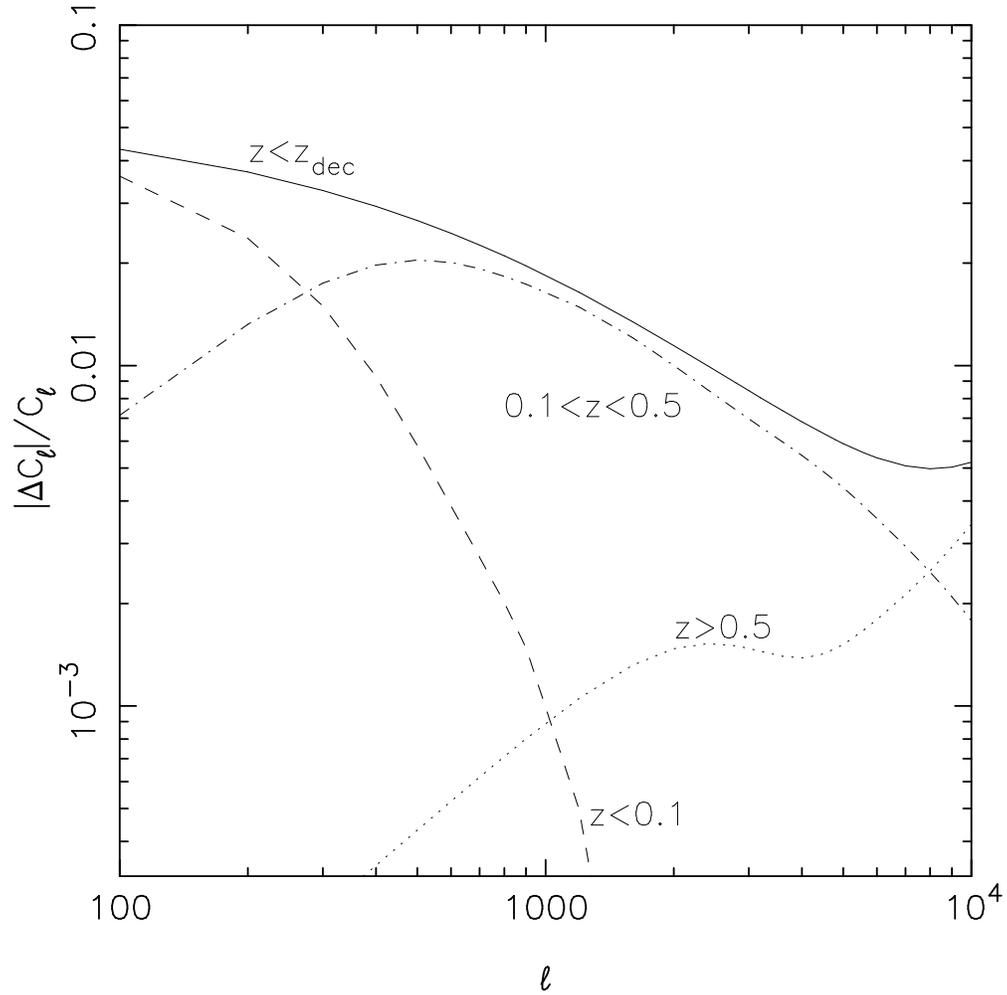}
\caption{Relative corrections of cluster radio halos to the 
thermal SZ angular power spectrum evaluated at $\nu=30$ GHz and 
for $f_{\rm RH} = 1/3$ (solid line). Contribution of clusters 
in different redshift ranges are also shown.  
\label{fig1}}
\end{figure}
\clearpage

\begin{figure}
\epsscale{0.8}   
\plotone{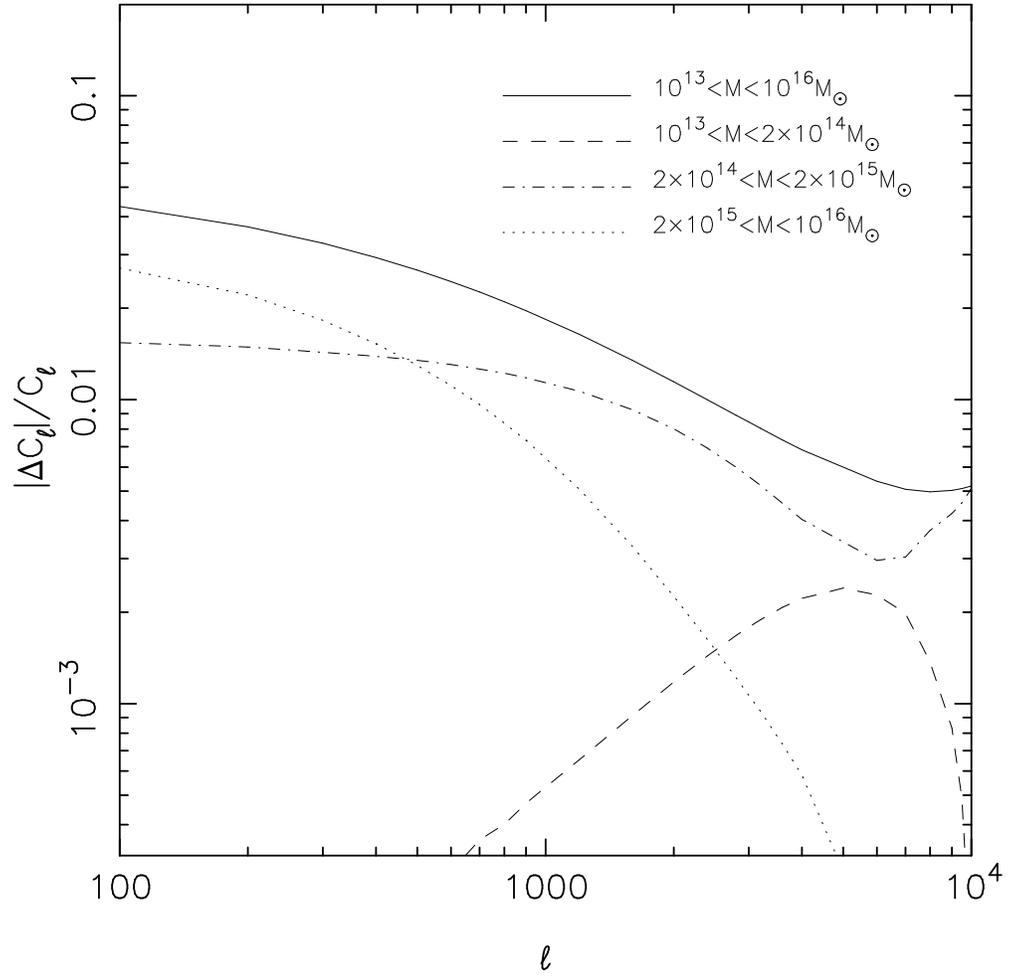}
\caption{The same as in Fig.1 but for different cluster mass ranges. 
\label{fig2}}
\end{figure}
\clearpage

\begin{figure}
\epsscale{0.8}   
\plotone{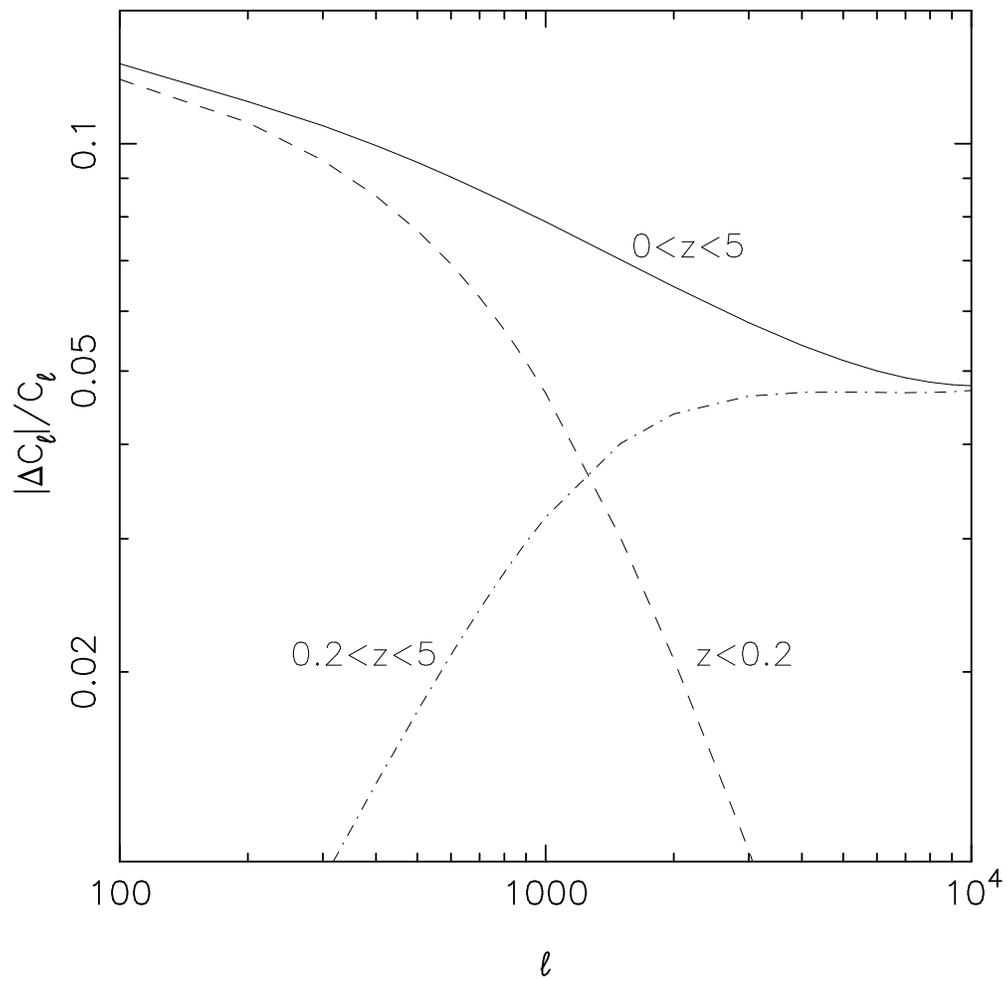}
\caption{The same as in Fig.1 but for cluster radio galaxies. 
\label{fig3}}
\end{figure}

\end{document}